\documentclass[reprint, nofootinbib, superscriptaddress]{revtex4-1}
\usepackage[utf8]{inputenc}
\usepackage{bm, amsthm, amsmath, amsfonts, amssymb, color, graphicx, natbib}%\usepackage{enumerate}
\usepackage{slashed}
\usepackage{xcolor}
\usepackage{verbatim}
\usepackage{float}
\usepackage[utf8]{inputenc}
\usepackage{hyperref}
\hypersetup{
	colorlinks=true,
	linkcolor=blue,
	filecolor=blue,      
	urlcolor=blue,
	citecolor=blue
}

\begin{document}

\title{Detectability of Primordial Black Hole Binaries at High Redshift}

\author{Qianhang Ding}
\email{qdingab@connect.ust.hk}
\affiliation{Department of Physics, The Hong Kong University of Science and Technology, Hong Kong, P.R.China}
\affiliation{Jockey Club Institute for Advanced Study, The Hong Kong University of Science and Technology, Hong Kong, P.R. China}

\begin{abstract}
	We show that the gravitational wave signals from primordial black hole (PBH) binaries at high redshift can be detected. The detectability of PBH binaries is enhanced by redshift bias and more PBH binaries at high redshift.  The initial clustering of PBHs is also included and enhances the effectively detectable mass ranges of PBHs at high redshift. Future observations on the gravitational wave at high redshift by space-based detectors such as LISA and SKA can constrain the fraction of PBHs in dark matter and PBHs initial distribution.
\end{abstract}

\maketitle
\section{Introduction}
Primordial black holes (PBHs) were born from primordial perturbations in highly overdense regions by gravitational collapse \cite{Hawking:1971ei, Dovich:1966no, Carr:1974nx}. Among modern mechanisms of PBHs formation, the inflationary origin \cite{Yokoyama:1995ex, Yokoyama:1998pt, Kawasaki:2016pql, Inomata:2016rbd, Cai:2018tuh} tightly connects the PBHs properties such as mass distribution \cite{Young:2014ana}, abundance distribution \cite{Belotsky:2014kca, Belotsky:2018wph, Bringmann:2018mxj,Ding:2019tjk}, the fraction of PBHs in dark matter \cite{Carr:2016drx} with the early universe evolution. Thus, the observational constraint on PBHs provides important clues for inflationary model building. 

PBHs play an important role in understanding the early universe. Directly observing signals from PBHs is the key step. In order to achieve that, distinguishing the PBHs and astrophysical black holes is essential. The method to distinguish black holes comes from the black holes 
intrinsic properties include mass, spin, charge and spacetime properties such as the spatial distribution and redshift.

One way to distinguish black holes comes from the mass of black holes. The mass of astrophysical black holes is heavier than a particular mass (around $3$ solar masses \cite{Rhoades:1974fn, Kalogera:1996ci}). Some specific mechanisms can produce around $1$ solar mass astrophysical black holes, see \cite{Kouvaris:2018wnh, Dasgupta:2020mqg}. However, PBHs have a different mass range. The mass of PBHs can be roughly estimated as follows \cite{Carr:2016drx},
\begin{align}\label{pbhmass}
	M \sim \frac{c^3 t}{G} \sim 5.03 \times 10^{4} \left(\frac{t}{1 \mathrm{s}}\right) M_{\odot}~,
\end{align}
where $t$ denotes the time when the corresponding perturbation returns to the horizon. Eq.~\eqref{pbhmass} shows that PBHs born before $\mathcal{O}(10^{-4})$ second since the hot big bang have mass less than $1$ solar mass, which is less than the minimal mass of astrophysical black holes. Considering Hawking radiation \cite{Hawking:1974rv}, PBHs with an initial mass of less than $10^{15} \mathrm{g}$ have already evaporated by now, as a result, PBHs with an initial mass of more than $10^{15} \mathrm{g}$ and less than $1$ solar mass still exist and produce various distinct signals in the present universe. In detecting sub-solar-mass PBHs, microlensing effect \cite{Paczynski:1985jf, Alcock:1993eu, Aubourg:1993wb, Alcock:2000ph}, femtolensing effect \cite{Barnacka:2012bm}, disruption of white dwarfs \cite{Graham:2015apa}, disruption of neutron stars \cite{Capela:2013yf, Kouvaris:2013kra, Pani:2014rca} and so on put constraints on the fraction of sub-solar-mass PBHs in dark matter. Some interactions between black holes and stars can help detect sub-solar-mass PBHs such as superradiance instability around black holes \cite{Baumann:2019ztm, Ding:2020bnl}. However, these observational effects such as microlensing are extremely weak, only a few events give a weak constraint on the fraction of PBHs in the dark matter.  

The other way to distinguish black holes depends on the redshift of signals from black holes. The astrophysical black holes forms after the death of Population III stars which are the first generation stars. Their observational signals exist at $z < 20$ \cite{Yoshida:2003ab}. In contrast, the redshift of PBHs goes through the whole history of the universe after the big bang. From this point, the signals from black holes with $z > 20$ should come from the PBHs. In detecting high redshift signals, gravitational wave is a powerful detection channel which carries information to travel through cosmological distance due to weak interaction between gravity and matter. Since the first gravitational wave event from the merger of binary black holes \cite{Abbott:2016blz}, the binary black holes system is the important gravitational wave source, so the detection of gravitational waves from PBH binaries is significant, which can bring us high redshift signals.

Recently, stellar mass PBHs have received renewed interest \cite{Bird:2016dcv, Clesse:2016vqa, Sasaki:2016jop} since the gravitational wave events detected by LIGO \cite{Abbott:2007kv} and Virgo \cite{TheVirgo:2014hva}. Intermediate-mass black holes show clues in LIGO and Virgo detection \cite{Abbott:2020tfl}. A supermassive black hole detection is being discussed \cite{Arzoumanian:2020vkk}. We propose to search for high redshift gravitational wave signals from PBH binaries, which covers a large mass range of PBHs at high redshift. Due to the  frequency of gravitational wave from high redshift source is highly redshifted, the usage of high frequency ground-based gravitational wave detector is limited. Targeting high redshift signals needs the low frequency space-based gravitational wave detector such as LISA \cite{LISA}, DECIGO \cite{Kawamura:2006up}, SKA \cite{dewdney2009square}. The detectability of PBHs shows the possibility in gravitational wave astronomy. There are some existing studies \cite{Nakamura:2016hna, Mandic:2016lcn} about gravitational wave from high redshift PBH binaries. In addition to their studies, we present the potential detectable mass range of PBH binaries at high redshift and include the effect of initial abundance distribution of PBHs on detectable mass range.

This paper is organized as follows. In Sect.~\ref{sensitivity}, we show that sensitivity of gravitational wave detectors can detect the high redshift gravitational wave from binary black holes. In Sect.~\ref{sec:eventrate}, we estimate the event rate of detectable signals from PBH binaries at high redshift.  In Sect.~\ref{effective}, we combine the gravitational wave detector sensitivity and PBH binaries event rate to obtain the effectively detectable mass range for PBH binaries at high redshift. We also discuss potential improvements in detectability by the initial clustering of PBHs.

\section{GW sensitivity at high redshift}\label{sensitivity}
The sensitivity of gravitational wave detectors is determined by the amplitude of gravitational wave and noise strain of the detector. The amplitude of gravitational wave from binary system can be expressed as
\begin{align}\label{amplitude}
	h = \frac{4}{d_L(z)} \bigg(\frac{G \mathcal{M}_{c}(1+z)}{c^2}\bigg)^{5/3} \bigg(\frac{\pi f}{c}\bigg)^{2/3}~.
\end{align} 
Here, $d_{L}(z)$ is the luminosity distance of gravitational wave source in flat FRW universe, which can be calculated from $d_{L}(z) = (1 + z) \int_0^z c /H(z') dz'$, $G$ is the Newton's constant, $c$ is the speed of light, $\mathcal{M}_{c}$ is the chirp mass of the binary system, which is defined as $\mathcal{M}_{c} \equiv (m_1 m_2)^{3/5}/(m_1 + m_2)^{1/5}$, and $f$ is the frequency of the gravitational wave. 

\begin{figure}[htbp] \centering
	\includegraphics[width=8cm]{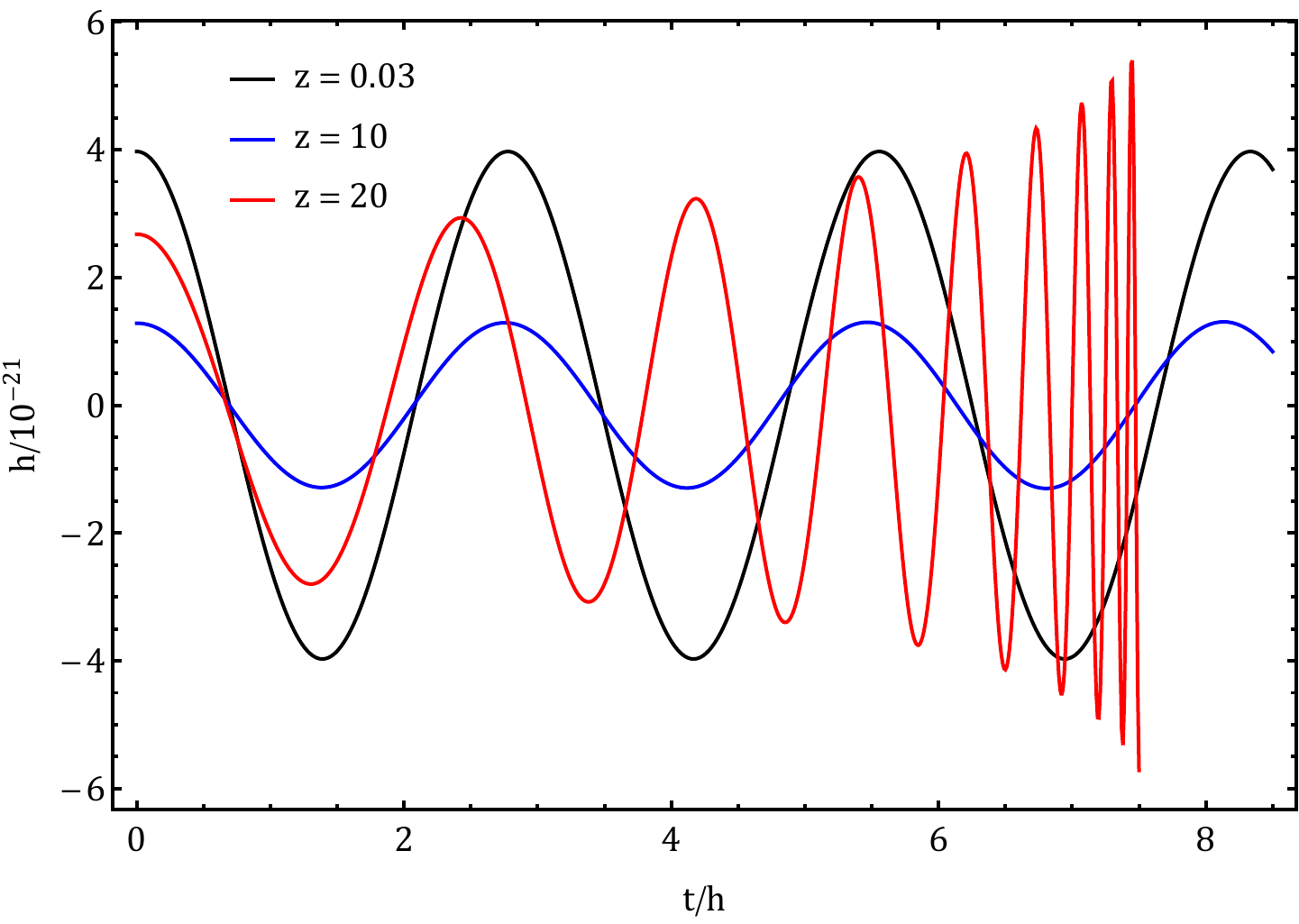}
	\caption{\label{fig:zbias}
		Gravitational waves with observed frequency $10^{-4} \mathrm{Hz}$ at redshift $z = 0.03$, $10$, $20$, also see in \cite{Rosado:2015voo}. The signals at high redshift stay later stage in binary evolution whose frequency is $f_{\mathrm{rest}} = (1 + z) f_{\mathrm{obs}}$, which increase the amplitude for observation.
	}
\end{figure}

The amplitude of the gravitational wave can be boosted by redshifted chirp mass $(1+z)\mathcal{M}_c$, which introduces an effect called redshift bias \cite{Rosado:2015voo}. Redshift bias of rest frame chirp mass can improve the detectability of binary system at high redshift. It works as follows. The amplitude and frequency of gravitational waves increase with time. The gravitational wave with higher frequency stays at a later stage in the inspiral phase, which has a larger amplitude than those with low frequency. Then, the frequency of high redshift gravitational waves is redshifted to the observable frequency at present which follows $f_{\mathrm{ rest}} = (1 + z) f_{\mathrm{obs}}$. The amplitude of high redshift gravitational wave also decreases because of large luminosity distance $d_{L}(z)$. Consequently, at observable frequency band, amplitude of high redshift gravitational wave depends on two factors: later stage in the inspiral phase and $d_{L}(z)$, where amplitude firstly increases and then decreases, which implies amplitude of gravitational wave doesn't always decrease monotonically with redshift, it has a minimum value at $z_{\mathrm{min}}$. In $\Lambda \mathrm{CDM}$ model, $z_{\mathrm{min}} = 2.63$ \cite{Rosado:2015voo}. Redshift bias can effectively improve the detectability of gravitational wave at high redshift. This effect is shown in Fig.~\ref{fig:zbias}.

In order to detect a gravitational wave event with a high confidence level, the signal-to-noise ratio (SNR) should exceed a threshold during observation. We follow  \cite{Rosado:2015voo}, where the optimal SNR is defined as
\begin{align}\label{snr}
	\mathrm{S/N} = \sqrt{4 \int_{f_{\mathrm{min}}}^{f_{\mathrm{max}}} \frac{|\tilde{h}(f)|^2}{S_n(f)} df}~.
\end{align}
Here, $\tilde{h}(f)$ is the Fourier transform of $h(t)$ based on the stationary phase approximation \cite{Droz:1999qx}, which is
\begin{align}\label{fourier}
	\tilde{h}(f) = \sqrt{\frac{5}{24}} \frac{(G \mathcal{M}_c (1+z))^{5/6}}{\pi^{2/3} c^{3/2} d_{L}(z)} f^{-7/6}~.
\end{align} 
The $S_n(f)$ is the noise strain of the detector. The $f_{\mathrm{min}}$ is the initial observed frequency and $f_{\mathrm{max}}$ is the final frequency during the observation time, which can be calculated from the evolution of observed gravitational wave frequency
\begin{align}\label{df}
	\frac{df}{dt} = \frac{96 \pi^{8/3} (G \mathcal{M}_c (1+z))^{5/3}}{5 c^5} f(t)^{11/3}~.
\end{align}
The optimal signal-to-noise ratio should exceed a conservative threshold of $8$, which makes sure the detection probability $> 95 \%$ and a false alarm probability $< 0.1\%$~.

Due to the redshift bias effect, $\mathrm{S/N}$ doesn't always decrease monotonically with redshift. It also has a minimal value at $z_{\mathrm{min}}$. When $z < z_{\mathrm{min}}$, optimal SNR decreases with $z$. Otherwise, optimal SNR increases with $z$ for the case $z > z_{\mathrm{min}}$. In order to obtain $z_{\mathrm{min}}$, we need calculate $\partial (\mathrm{S/N})^2 / \partial z = 0$, which can be expressed as following,
\begin{align}\label{zmineq}
	\frac{\partial \mathrm{S/N}^2}{\partial z} = 4 \int_{t_{\mathrm{min}}}^{t_{\mathrm{max}}} \frac{1}{S_n(f)} \frac{\partial}{\partial z}(|\tilde{h}(f)|^2 \frac{df}{dt}) dt = 0~.
\end{align}
Applying Eq.~\eqref{fourier} and \eqref{df} into Eq.~\eqref{zmineq}, we can get equation for $z_{\mathrm{min}}$,
\begin{align}\label{zmineq_result}
	(1 + z_{\mathrm{min}}) \frac{\partial \ln d_{L}(z)}{\partial z} \bigg|_{z_\mathrm{min}} = \frac{5}{3}~.
\end{align}
In $\Lambda \mathrm{CDM}$ model with $\Omega_{\mathrm{m}} = 0.315$ and $\Omega_{\Lambda} = 0.685$ \cite{Aghanim:2018eyx}, $z_\mathrm{min} = 2.63$. The Fig.~2 of \cite{Rosado:2015voo} shows the same result.

\begin{figure}[htbp] \centering
	\includegraphics[width=8cm]{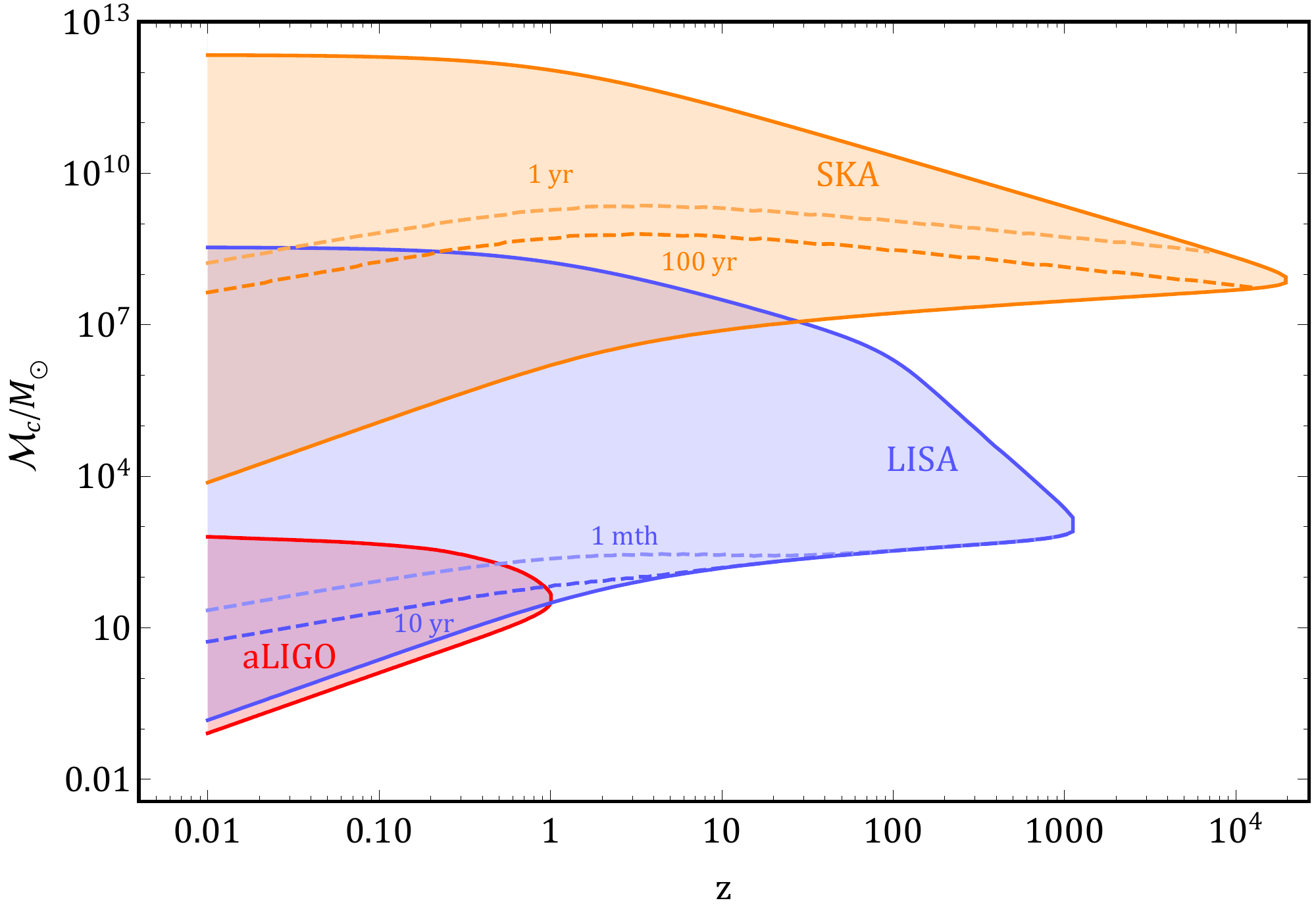}
	\caption{\label{fig:masszrange}
		The detectable mass range at different redshifts for gravitational wave detectors aLIGO, LISA and SKA .The SNR of each binary ($\mathcal{M}_c$, $z$) is larger than the threshold of $8$. The typical observation time $1 \mathrm{month}$ and $10 \mathrm{year}$ for LISA, $1 \mathrm{year}$ and $100 \mathrm{year}$ for SKA have been shown as dashed lines, where $\mathrm{S/N} = 8$. The shadow regions above dashed lines show the detectable mass ranges within different observation time. }
\end{figure}

Combine Eqs.~\eqref{snr}\eqref{fourier}\eqref{df} with the condition $\mathrm{S/N} > 8$, we can get detectable mass ranges at different redshifts in Fig.~\ref{fig:masszrange}.
In order to observe the mass range as large as possible, the frequency range in Eq.~\eqref{snr} between $f_{\mathrm{min}}$ and $f_{\mathrm{max}}$ should be large so that $\mathrm{S/N} > 8$. We set the $f_{\mathrm{min}}$ as the minimal detectable frequency of the detector. The $f_{\mathrm{max}}$ is set to the minimal quantity between the maximal detectable frequency of the detector and the maximal redshifted inspiral phase frequency $f_{\mathrm{insp}}^{\mathrm{max}}/(1+z) \mathrm{Hz}$. Here, the maximal inspiral phase frequency in rest frame can be estimated as $f_{\mathrm{insp}}^{\mathrm{max}} = (a \eta^2 + b \eta + c)/\pi G M$ \cite{Chernoff:1993th, Ajith:2009bn, Zhu:2011bd}. The symmetric mass ratio $\eta$ is defined as $\eta \equiv m_1 m_2/M^2$, $M \equiv m_1 + m_2$ and the coefficients $a=0.29740$, $b=0.04481$, $c=0.09556$ (See Table 1 of \cite{Ajith:2007kx}). The $S_n(f)$ for each detector can be obtained from Fig.~A2 in \cite{Moore:2014lga}.

In Fig.~\ref{fig:masszrange}, there is a redshift range corresponding with a PBH mass. For low mass PBH binaries, amplitude of gravitational wave is small, which needs a higher $f_{\mathrm{max}}$ to increase SNR, then the maximal detectable redshift should be small so that $f_{\mathrm{max}}$ isn't compressed by $f_{\mathrm{insp}}^{\mathrm{max}}/(1+z) \mathrm{Hz}$. The boundary of shadow region corresponds with $\mathrm{S/N} = 8$. The PBH binary at higher redshift can not be detected due to the condition $f_{\mathrm{min}} \geq f_{\mathrm{max}}$ in Eq.~\eqref{snr}, which results from the maximal inspiral frequency of binary system is redshifted that is less than the minimal frequency of the detector. For a practical observation, the observation time $T_{\mathrm{obs}}$ is limited, so we choose different observation time to determine the frequency range as following,
\begin{align}\label{frerange}
	 f_{\mathrm{max}} - f_{\mathrm{min}} = \int_{t_\mathrm{ini}}^{t_{\mathrm{ini}} + T_{\mathrm{obs}}} \frac{df}{dt} dt~.
\end{align}
The $f_{\mathrm{min}}$ is chosen from the frequency band of detector, then $f_{\mathrm{max}}$ can be obtained from Eq.~\eqref{frerange}. The typical observation time has been shown as dashed line in Fig.~\ref{fig:masszrange}. The shadow regions above dashed lines show detectable mass ranges of detectors within the observation time. We can find dashed lines are not monotonic, which is caused by the redshift bias effect. According to Eq.~\eqref{zmineq} and \eqref{zmineq_result}, optimal SNR function is monotonically decreasing for $z < 2.63$ and increasing for $z > 2.63$ in our $\Lambda \mathrm{CDM}$ universe \cite{Aghanim:2018eyx}, for detecting the signal with $\mathrm{S/N} = 8$, the chirp mass of PBH binary need increase for $z < 2.63$ and decrease for $z > 2.63$.

\section{Event rate of PBH binaries}\label{sec:eventrate}
In order to observe the PBHs at high redshift, not only should gravitational wave sensitivity be high but also the event rate of PBH binaries system should be large enough. The estimation of PBH binaries event rate follows \cite{Sasaki:2016jop}, the condition for decouple of PBH binaries from FRW background at matter-radiation equality $z = z_{\mathrm{eq}}$ is
\begin{align}
	x < f^{1/3}_{\mathrm{PBH}} \bar{x}~.
\end{align}
Here, $x$ is the physical distance between two nearby black holes, $f_{\mathrm{PBH}}$ is the fraction of PBHs in the dark matter. $\bar{x}$ is the mean separation of PBHs at matter-radiation equality, which is determined by PBH mass $M_{\mathrm{BH}}$ and energy density $\rho_{\mathrm{BH}}(z_{\mathrm{eq}})$ as $\bar{x} = (M_{\mathrm{BH}}/\rho_{\mathrm{BH}}(z_{\mathrm{eq}}))^{1/3}$. After interacting with the third black hole, the probability distribution function of binary system with respect to the major axes $a$ and eccentricity $e$ \cite{Sasaki:2016jop} is
\begin{align}\label{pae}
	P(a, e) = \frac{3}{4} \frac{f^{3/2}_{\mathrm{PBH}} a^{1/2}}{\bar{x}^{3/2}} \frac{e}{(1-e^2)^{3/2}}~.
\end{align}
After long time evolution, the major axes shrinks to a suitable quantity that the frequency of gravitational wave lies in the frequency band of detector. Following \cite{Peters:1963ux, Peters:1964zz}, we assume that $e$ keep the constant during inspiral phase, the later major axes can be expressed as
\begin{align}\label{majora}
	a^4 = a_{\mathrm{ini}}^4 - \frac{256}{5}\frac{G^3 \mu  M^2}{c^5 (1-e^2)^{7/2}} t~,
\end{align}
where $a_{\mathrm{ini}}$ is the initial major axes of PBH binary, $t$ is the evolution time, $M \equiv m_1+m_2$, $\mu \equiv m_1 m_2/M$. Applying the Kepler's third law with Eq.~\eqref{majora}, Eq.~\eqref{pae} can be written into
\begin{align}
	P(t, e) = \frac{48}{5}\frac{f^{3/2}_{\mathrm{PBH}} G^3 \mu  M^2}{c^5 \bar{x}^{3/2} T^{5/8}} \frac{e}{(1-e^2)^5}~.
\end{align}
Here, $T$ is defined as 
\begin{align}
T \equiv \left(\frac{G M}{\pi^2 f_{GW}^2}\right)^{4/3} + \frac{256}{5}\frac{G^3 \mu M^2}{c^5 (1 - e^2)^{7/2}} t~.
\end{align}
$f_{GW}$ is the frequency of gravitational waves. Then, integrate the $e$ from $0$ to the maximum eccentricity to get a probability distribution function of time $P(t)$, the maximum eccentricity is determined by binary formation conditions in \cite{Sasaki:2016jop}.

\begin{figure}[htbp] \centering
	\includegraphics[width=8cm]{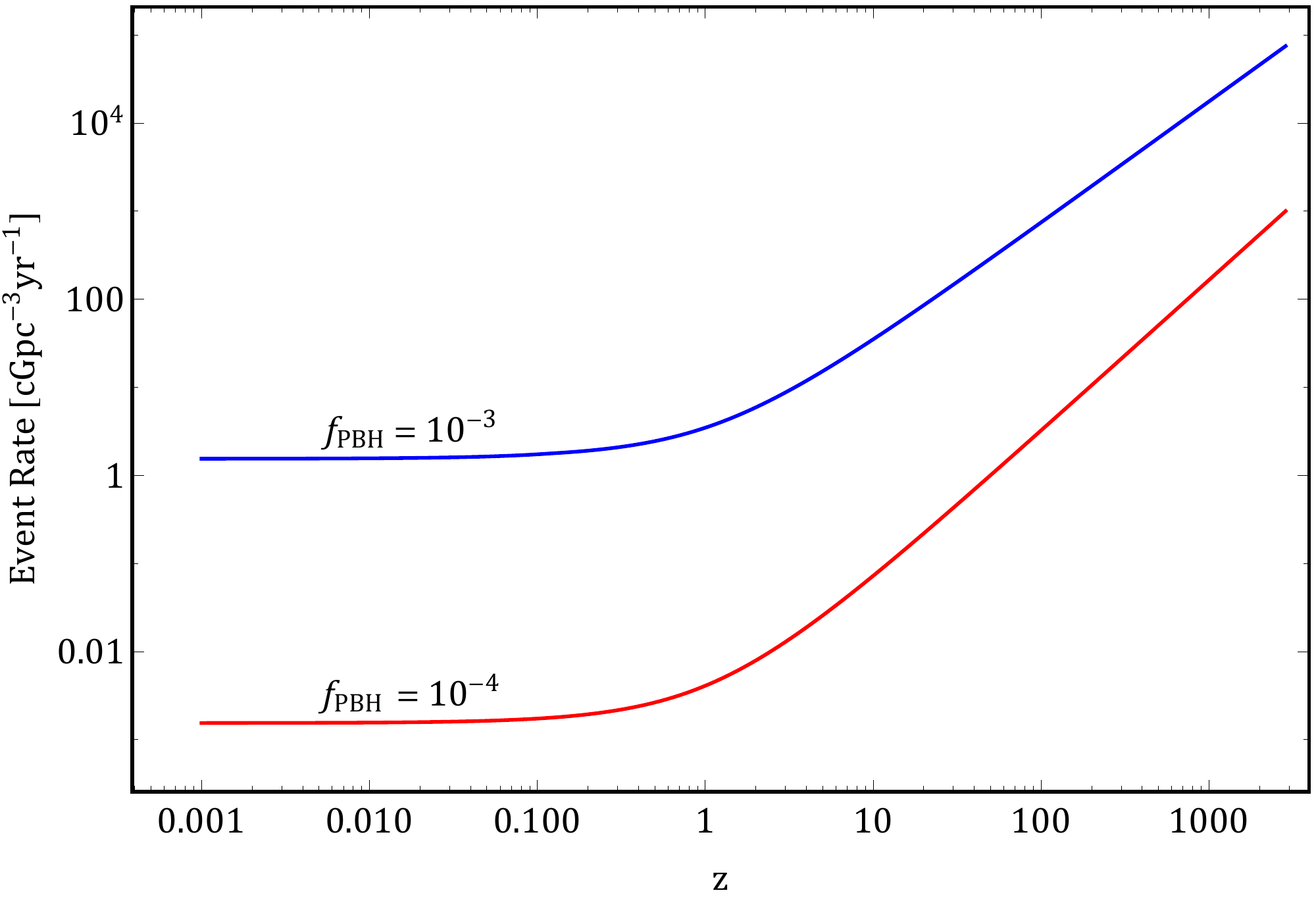}
	\caption{\label{fig:eventrate}
		The event rate for PBH binaries with $100$ solar masses and the observed frequency is $10^{-4} \mathrm{Hz}$ at different redshifts.	The event rate is measured in comoving volume with the unit of $\mathrm{cGpc}^{-3}\mathrm{yr}^{-1}$.}
\end{figure}

The event rate of gravitational wave signals from PBH binaries can be calculated from $P(t)$ and average number density of PBHs in comoving volume $n_{\mathrm{BH}}$, 
\begin{align}\label{eventrate}
	\mathrm{event~rate} = n_{\mathrm{BH}} P(t) = \frac{3 H_0^2}{8 \pi G} \frac{\Omega_{\mathrm{BH}}}{M_{\mathrm{BH}}} P(t)~.
\end{align}
In order to obtain the event rate at different redshifts, the time-redshift relation along the line of sight is required, which can be calculated from,
\begin{align}\label{tz}
	t(z) = \int_{z}^{\infty}\frac{dz'}{H(z')(1+z')}~.
\end{align}
Applying Eq.~\eqref{tz} into $P(t)$. Then, we can get the event rate of PBH binary at different redshifts in Fig.~\ref{fig:eventrate}.
It shows the increasing event rate of PBH binaries at higher redshift. This is caused by $P(a) \propto a^{-1/4}$ \cite{Sasaki:2016jop}, there are more PBH binaries having small major axes, when binaries form, which emit high frequency gravitational wave and merge at high redshift. So more high frequency gravitational wave events propagate from high redshift to the present that are redshifted and lied in frequency band of detectors. (also see in \cite{Ali-Haimoud:2017rtz, Kavanagh:2018ggo}).

\begin{figure}[htbp] \centering
	\includegraphics[width=8cm]{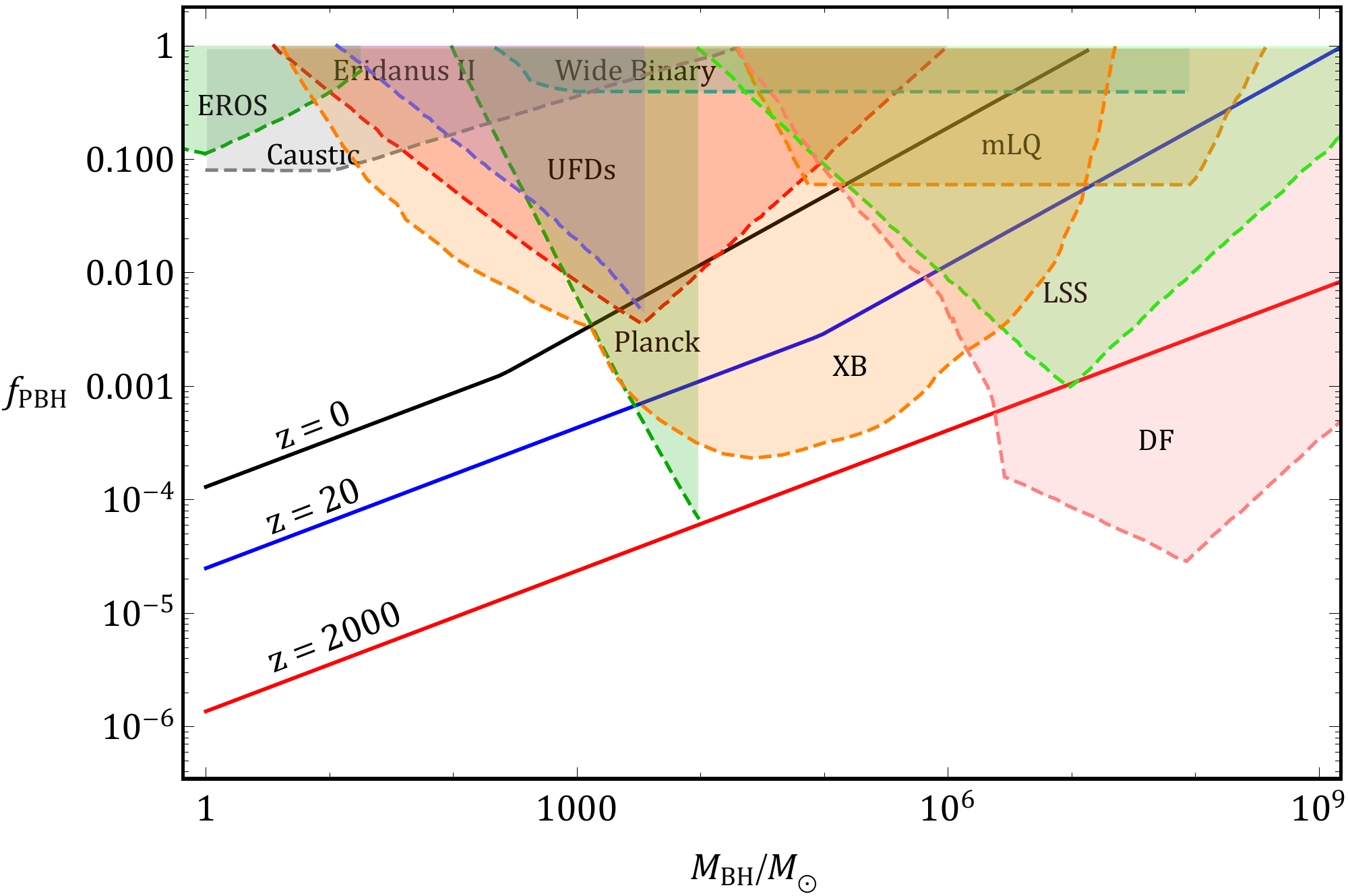}
	\caption{\label{fig:fpbh}
		The fraction of PBHs in dark matter with respect to PBH mass $M_{\mathrm{BH}}$ at different redshifts, to reach an event rate of $1 \mathrm{cGpc}^{-3} \mathrm{yr}^{-1}$. Limits from EROS \cite{Tisserand:2006zx}, Caustic \cite{Oguri:2017ock}, UFDs \cite{Brandt:2016aco}, Eridanus II \cite{Brandt:2016aco}, Plank \cite{Ali-Haimoud:2016mbv}, wide-binary disruption \cite{Quinn:2009zg}, accretion limits come from X-ray binaries \cite{Inoue:2017csr}, millilensing of quasars \cite{Wilkinson:2001vv},  generation of large-scale structure through Poisson fluctuations \cite{Afshordi:2003zb} and dynamical friction on halo objects \cite{Carr:1997cn} are also plotted}
\end{figure}

More event rates of PBH binaries at high redshift in Fig.~\ref{fig:eventrate} shows that it is possible to detect the signal from PBH binaries at high redshift, even though, the fraction of PBHs in dark matter $f_{\mathrm{PBH}}$ is strongly constrained by astronomy observation. Then, we can compare high redshift $f_{\mathrm{PBH}}-M_{\mathrm{BH}}$ relation with the astronomy observation. Here, we consider the monochromatic PBH mass function in Eq.~\eqref{eventrate}. Considering the detectable merger rate of binary black holes is $2-53 \mathrm{cGpc}^{-3} \mathrm{yr}^{-1}$ for aLIGO \cite{LIGOScientific:2016kwr}, a reasonable and detectable event rate is set to $1 \mathrm{cGpc}^{-3} \mathrm{yr}^{-1}$. Then we can get the relation among the fraction of PBHs in dark matter $f_{\mathrm{PBH}}$, PBH mass $M_{\mathrm{BH}}$ and redshift $z$ from Eq.~\eqref{eventrate}, which is shown in Fig.~\ref{fig:fpbh}.
Three typical redshift curves $z = 0,~20,~2000$ are shown, which corresponds with a fixed event rate $1 \mathrm{cGpc}^{-3} \mathrm{yr}^{-1}$. We can find that at the same $M_{\mathrm{BH}}$, the higher redshift, the lower $f_{\mathrm{PBH}}$ we need to reach an detectable event rate of $1 \mathrm{cGpc}^{-3} \mathrm{yr}^{-1}$, which means high redshift PBH binaries can be detected without ruling out by astronomy constraints. It results from that there are more event rates at high redshift, when we fix $f_{\mathrm{PBH}}$ as shown in Fig.~\ref{fig:eventrate}. Then we don't need a large $f_{\mathrm{PBH}}$ to produce an detectable event rate of $1 \mathrm{cGpc}^{-3} \mathrm{yr}^{-1}$ at high redshift.

\section{Detectable mass range at high redshift}\label{effective}
\begin{figure}[htbp] \centering
	\includegraphics[width=8cm]{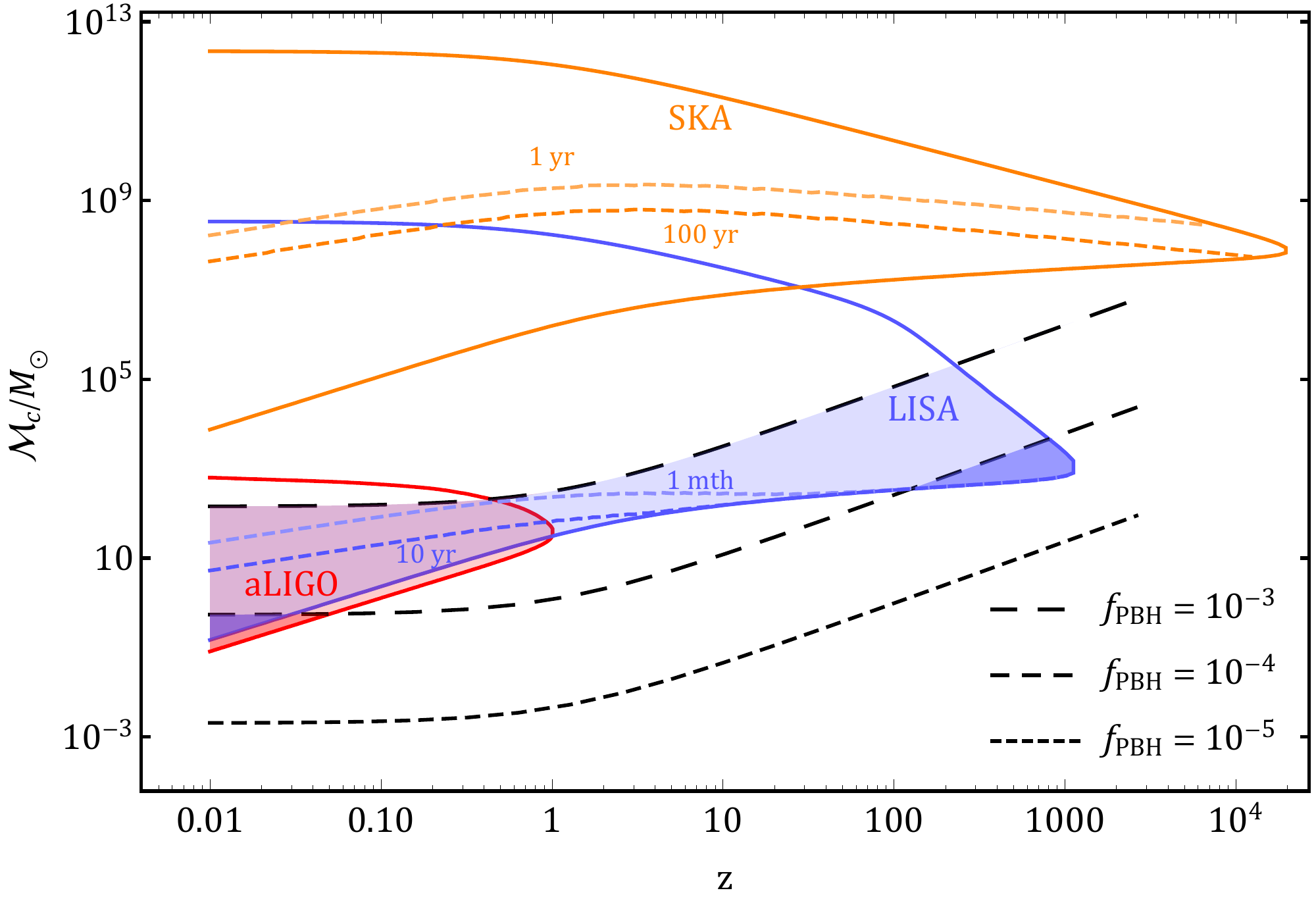}
	\caption{\label{fig:obs}
		The effectively detectable regions for PBH binaries at different redshifts. The solid lines give the boundary of parameter regions that detectors can observe. The black dashed lines give the critical mass with an event rate of $1 \mathrm{cGpc}^{-3} \mathrm{yr}^{-1}$ that we can observe for different $f_{\mathrm{PBH}}$. The shadow regions below the black dashed lines and above the solid lines are the effectively detectable regions for PBH binaries. }
\end{figure}

Above all, we have discussed the detectable mass ranges of detectors aLIGO, LISA, SKA and the event rate of PBH binary at different redshifts. Now, we can combine the two results to obtain the effectively detectable region for PBHs at different redshifts as shown in Fig.~\ref{fig:obs}.

In drawing the effective regions, we assume the effectively detectable range follows the conditions, 
\begin{align}\label{obscon}
	\mathrm{event~rate} > 1 \mathrm{cGpc}^{-3} \mathrm{yr}^{-1}~, ~~~ f_{\mathrm{PBH}} < f_{\mathrm{unc}}~.
\end{align}
Here, we follow \cite{Ding:2019tjk, LIGOScientific:2016kwr}, the minimal detectable event rate is set to $1 \mathrm{cGpc}^{-3} \mathrm{yr}^{-1}$. $f_{\mathrm{unc}}$ is the unconstraint fraction of PBHs in the dark matter where we take $10^{-3}$, $10^{-4}$, $10^{-5}$ for illustration. The black dashed lines correspond with an event rate of $1 \mathrm{cGpc}^{-3} \mathrm{yr}^{-1}$. The shadow regions show the effectively detectable mass ranges at high redshift. One thing we need to pay attention to is the maximum detectable redshift is $z = 3000$, which comes from the PBH binary model \cite{Sasaki:2016jop}, the PBH binaries formed at matter-radiation equality $z_{\mathrm{eq}} = 3000$. This mechanism can hardly produce PBHs in radiation-domination epoch at $z > z_{\mathrm{eq}}$, which makes it difficult in observing a larger mass range and detecting PBH binaries on SKA.

\begin{figure}[htbp] \centering
	\includegraphics[width=8cm]{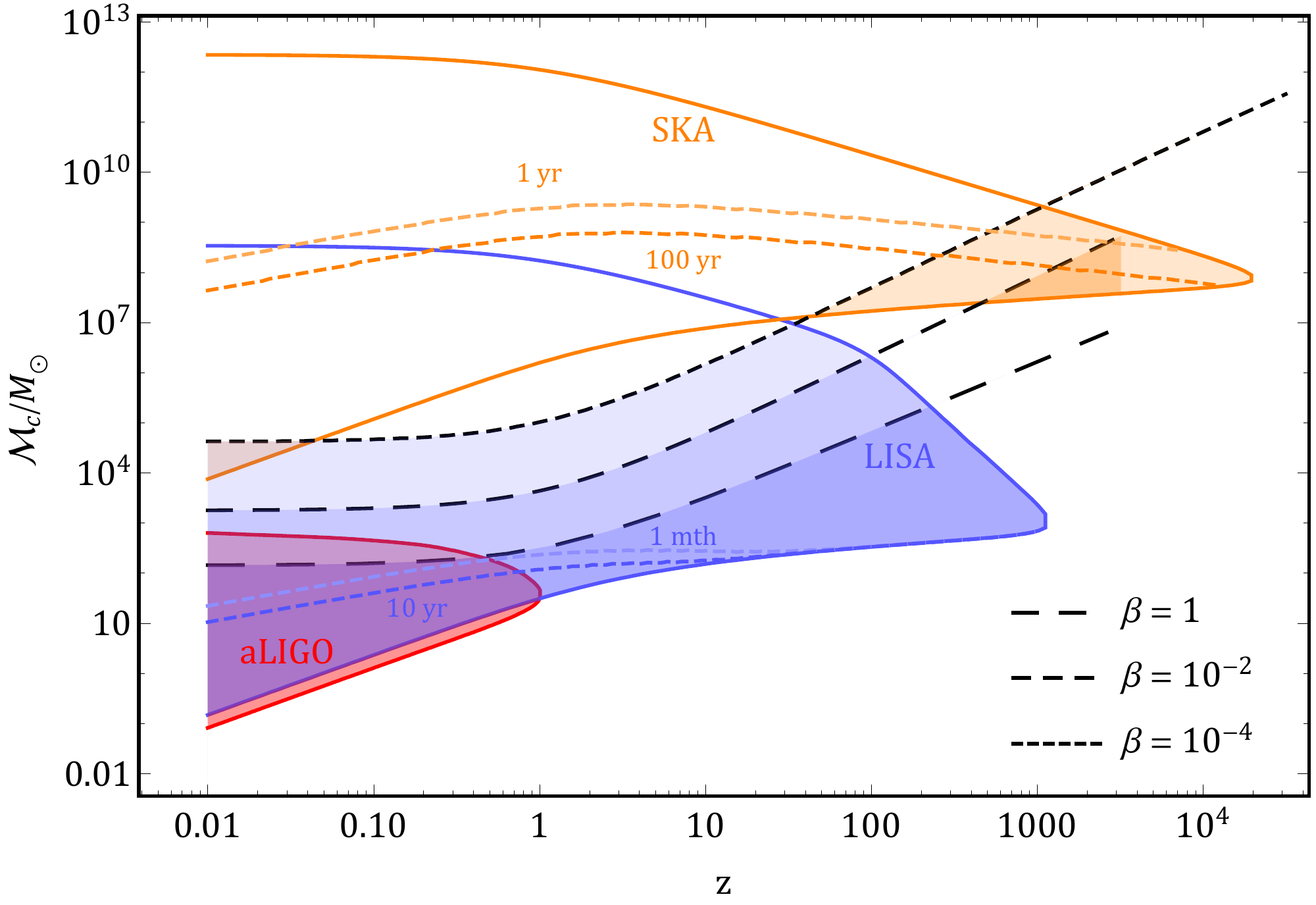}
	\caption{\label{fig:clustering}
		The effectively detectable regions for PBH binaries at different redshifts with initial clustering. The solid lines give the boundary of parameter regions that detectors can observe. The black dashed lines give the critical mass with an event rate of $1 \mathrm{cGpc}^{-3} \mathrm{yr}^{-1}$ and $f_{\mathrm{PBH}} = 10^{-3}$ that we can observe for different $\beta$. The shadow regions below the black dashed lines and above the solid lines are the effectively detectable regions for PBH binaries. }
\end{figure}

The inclusion of initial spatial clustering of PBHs changes the story. The initial clustering of PBHs forms the high density PBHs regions in radiation-domination epoch. As we have shown in \cite{Ding:2019tjk}, the initial clustering of PBHs can be constructed in multi-stream inflation \cite{Li:2009sp, Li:2009me, Wang:2010rs, Afshordi:2010wn}. During multi-stream inflation, inflaton travels along two inflation trajectories, inflation potential on one trajectories can produce PBHs, the others can not produce PBHs. In the PBHs clustering model \cite{Ding:2019tjk}, the PBHs clustering regions are denoted by $\beta$, which is the volume fraction of PBHs regions in the observed universe. The redshift of matter-radiation equality in PBHs rich regions can be calculated by increasing the PBHs fraction in dark matter in PBHs rich region $f_{\mathrm{PBH}} \rightarrow f_{\mathrm{PBH}}/\beta$ as following
\begin{align}
	1 + z_{\mathrm{PBH}, \mathrm{eq}} = (1 - f_{\mathrm{PBH}} + \frac{f_{\mathrm{PBH}}}{\beta}) (1 + z_{\mathrm{eq}})~.
\end{align}
The high density PBHs region improves the probability of PBH binaries formation by increasing the PBHs density, then event rate of PBH binaries can be enhanced due to a larger PBH binaries density, which results in that massive PBH binaries can meet the Eq.~\eqref{obscon}, the result is shown in Fig.~\ref{fig:clustering}.
The initial clustering of PBHs expands the detectable mass range for PBH binaries and improves the detectability of PBH binaries in SKA. However, the clustering volume fraction $\beta$ cannot be infinitesimal, this could result in few PBHs surviving at present and most of them merged at very high redshift which leaves weak signals to detectors.

\section{Conclusion}
In this paper, we show that the PBH binaries can be detected by the high redshift $z > 20$ gravitational wave signals. During the propagation of gravitational wave from high redshift to the present, the frequency of gravitational wave is decreasing so that lies in the frequency band of gravitational wave detectors. Then amplitude of redshifted high frequency gravitational wave can significantly enhance the $\mathrm{S/N}$ at $z > z_{\mathrm{min}}$, where $z_{\mathrm{min}} = 2.63$ in $\Lambda \mathrm{CDM}$ model. This phenomena is detailed studied in \cite{Rosado:2015voo}. In order to detect gravitational wave signal from PBH binaries within practical observation time. The event rate of gravitational wave from PBH binaries is estimated. We consider the model in \cite{Sasaki:2016jop}, the PBH binaries events are enhanced at high redshift which can achieve $90 \mathrm{cGpc}^{-3}\mathrm{yr}^{-1}$ at $z = 20$ for PBHs mass $100$ solar masses with $f_{\mathrm{PBH}} = 10^{-3}$ in Fig.~\ref{fig:eventrate}. The enhancement of gravitational wave signals and the event rate from PBH binaries at high redshift improve the detectability of PBH binaries. Then, we study the effect of initial clustering of PBHs on its detectability as we have studied in \cite{Ding:2019tjk}. An enhancement in effectively detectable mass ranges appears by the initial clustering of PBHs. The effectively observable mass range covers $20 - 10^{3}$ solar masses at $z = 20$ within ten years observation time on LISA. This mass range can be enhanced to $20 - 10^6$ solar masses with initial clustering $\beta = 10^{-4}$. The wide mass range shows that it is possible to detect the PBH binaries on LISA and SKA at high redshift.

Based on the possibility of PBH binaries detection at high redshift, we can directly observe the signals from the PBHs. This can help us build an early universe model. The mass function of PBHs can be obtained from the chirp mass of gravitational wave in the rest frame and gives clues to the inflation potential \cite{Young:2014ana}. The abundance distribution of PBHs can be obtained from the position distribution of gravitational wave events, which gives hint for inflation model with position properties, such as the initial clustering of PBHs from multi-stream inflation \cite{Ding:2019tjk}. The fraction of PBHs in the dark matter can be constrained by the high redshift gravitational wave event rate, which can tell us the potential dark matter candidates.

In this work, we only consider the monochromatic PBH mass function. The extended mass function of PBHs can form extreme mass ratio inspiral, which can produce different signals. In detecting more PBHs events at high redshift, stochastic gravitational wave background from PBHs formation need to be deducted \cite{Wang:2016ana, Cai:2019jah, Nakama:2020kdc}. Due to the high redshift of the gravitational wave, no electromagnetic counterpart can be observed along with the gravitational wave signal. We cannot directly observe the luminosity distance of gravitational wave. However, the interaction between Population III stars with PBHs could be a potential source for electromagnetic signals at $z < 20$, which helps us understand PBHs evolution at high redshift. 

\section*{Acknowledgements}
I thank Yi Wang for detailed suggestions, Lingfeng Li and Xi Tong for useful comments. I would like to thank Xingwei Tang for her kindly encouragement and suggestions.

\end{document}